\documentclass{icrc2009}

\usepackage{graphicx}   
\usepackage[caption=false]{caption}    
\usepackage[font=footnotesize]{subfig} 
\usepackage{fixltx2e}
\usepackage{url}

\newcommand{\shorttitle}[1]%
{\markboth{Proceedings of the 31\MakeLowercase{$^{st}$} ICRC, {\L}\'{o}d\'{z} 2009}{#1} }
\newcommand{\etal}{\MakeLowercase{\textit{et al. }}} 

\usepackage{amsmath}

\begin{document}
\title{The Anticoincidence Counter System of AMS-02}

\author{\IEEEauthorblockN{Ph. von Doetinchem\IEEEauthorrefmark{1},
			  Th. Kirn\IEEEauthorrefmark{1},
                          K. L\"{u}belsmeyer\IEEEauthorrefmark{1} and
                          St. Schael\IEEEauthorrefmark{1}}
			  \\
\IEEEauthorblockA{\IEEEauthorrefmark{1}I. Physics Institute B, RWTH Aachen University, Sommerfeldstr. 14, 52074 Aachen, Germany}
}

\shorttitle{Ph. von Doetinchem \etal The ACC of AMS-02}
\maketitle

\begin{abstract}
The AMS-02 experiment will be installed on the International Space Station at an altitude of about 400\,km in 2010 to measure for three years cosmic rays. The total acceptance including the electromagnetic calorimeter is 0.095\,m$^2$sr.

This work focuses on the anticoincidence counter system (ACC). The ACC is a single layer composed of 16 interlocking scintillator panels that surround the tracker inside the inner bore of the superconducting magnet. The ACC needs to detect particles that enter or exit the tracker through the sides with an efficiency of better than 99.99\,\%. This allows to reject particles that have not passed through all the subdetectors and may confuse the charge and momentum measurements which is important for an improvement of the antinuclei-measurements.

In 2007/2008 all subdetectors were integrated into the AMS-02 experiment and atmospheric muons were collected. These data were used to determine the ACC detection efficiency.
  \end{abstract}

\begin{IEEEkeywords}
AMS-02, Veto, Inefficiency
\end{IEEEkeywords}

\section{The AMS-02 Detector}

\begin{figure}
\centering
\includegraphics[width=1.0\linewidth]{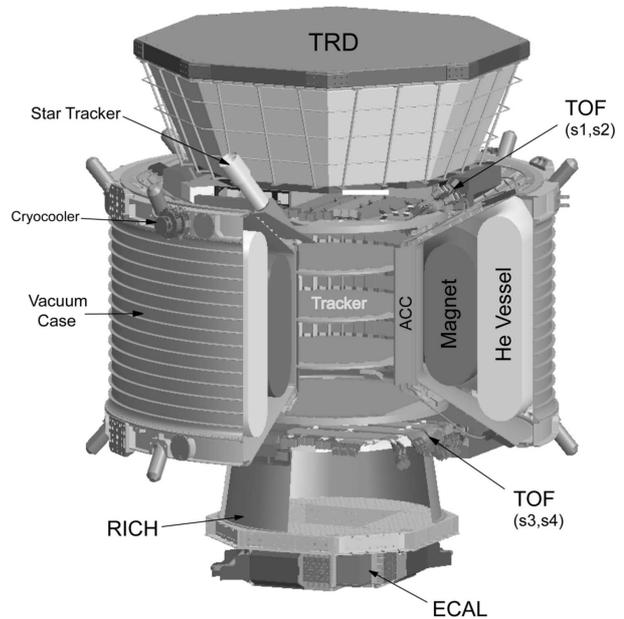}
\caption{The AMS-02 detector.}
\label{f-ams_detector}
\end{figure}

The AMS-02 experiment will be installed on the International Space Station at an altitude of about 400\,km in 2010 for about three years to measure cosmic rays without the influence of the Earth's atmosphere \cite{ams04}. The detector consists of several subdetectors for the determination of the particle properties, namely a transition radiation detector (TRD), a time of flight system (TOF), a cylindrical silicon microstrip tracker with eight layers surrounded by an anticoincidence counter system (ACC) in a superconducting magnet with a field of 0.8\,T strength, a ring image \v{C}erenkov detector (RICH) and an electromagnetic calorimeter (ECAL) (fig.~\ref{f-ams_detector}).

\section{The Anticoincidence Counter}

\begin{figure*}
\centering
\includegraphics[width=1.0\linewidth]{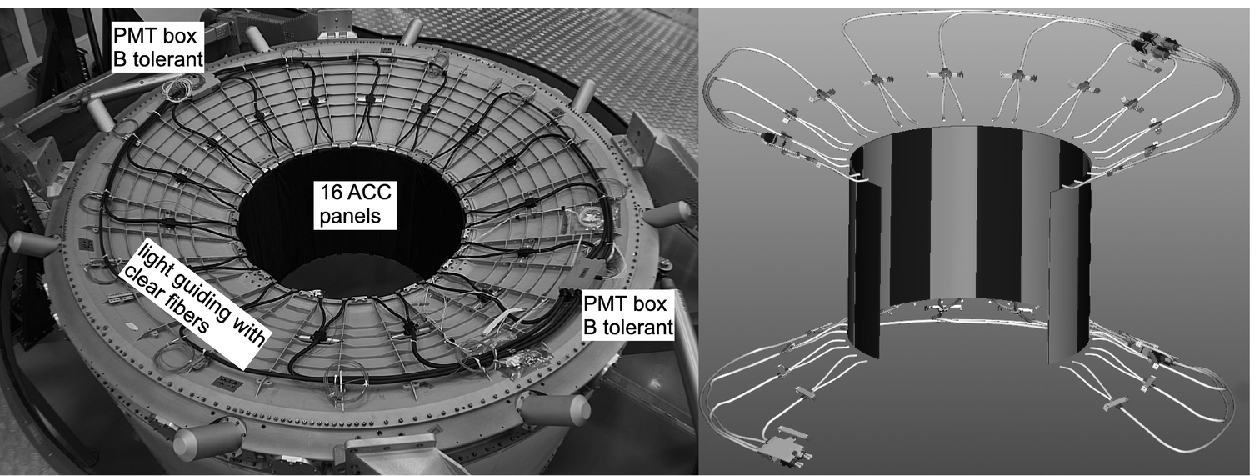}

\vspace{0.2cm}

\includegraphics[width=0.9\linewidth]{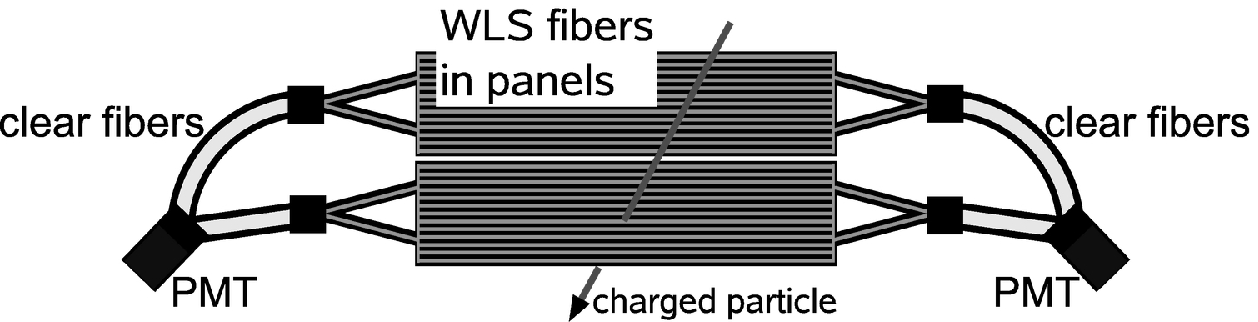}
\caption{\textbf{\textit{Upper)}} The anticoincidence counter system after integration (left) and the principle of component arrangement (right). \textbf{\textit{Lower)}} ACC working principle.}
\label{f-acc_overview}
\end{figure*}

The AMS-02 anticoincidence counter \cite{doe08,doe09} surrounds the silicon tracker and can be used as a veto for the trigger decision made by the TOF (fig.~\ref{f-acc_overview}, upper). This is important for rejecting events with particles entering the detector from the side or with particles from secondary interactions inside the detector which could distort the charge measurement. To improve existing upper limits on antihelium an inefficiency of the ACC smaller than $10^{-4}$ is needed according to MC simulations. The inefficiency is the ratio of missed to the total number of particle tracks crossing the ACC.

Another important task of the ACC is to reduce the trigger rate during periods of very large flux, e.g. in the South Atlantic Anomaly. For that purpose, it is important to use a detector with a fast response. 

The ACC cylinder has a diameter of 1.1\,m and a height of 0.83\,m and is made out of 16 scintillation panels (Bicron BC-414) with a thickness of 8\,mm. The ultraviolet scintillation light through ionization losses of charged particles is absorbed by wavelength shifting fibers (WLS, Kuraray Y-11(200)M) which are embedded into the panels. The WLS fibers are coupled to clear fiber cables (Toray PJU-FB1000) for the final light transport to photomultiplier tubes (Hamamatsu R5946). A set of two panels is read out by the same two photomultipliers, one on top and one on the bottom, via clear fiber cables (Y-shape) in order to have redundancy and to save weight (fig.~\ref{f-acc_overview}, lower). The slot between two panels is realized with tongue and groove and is crucial for the determination of the inefficiency because of less scintillator material and a smaller active (WLS) to passive (scintillator) material ratio. The qualification of the individual components is discussed in references~\cite{doe08,doe09}.

\section{ACC Inefficiency Determination with atmospheric Muons}

AMS-02 was pre-integrated with all subsystems but the magnet in 2007 and 2008 at CERN, Geneva. The data used here are taken after this installation. 

In most runs a trigger is given by any two out of the four TOF planes. The trigger condition for events used in this analysis used tracks with hits in both upper TOF layers. TRD and tracker tracks are extrapolated to the ACC in order to find the intersection on the ACC cylinder and to determine the ACC detection efficiency. 

\subsection{Event Reconstruction}

\begin{figure}
\includegraphics[width=1.0\linewidth]{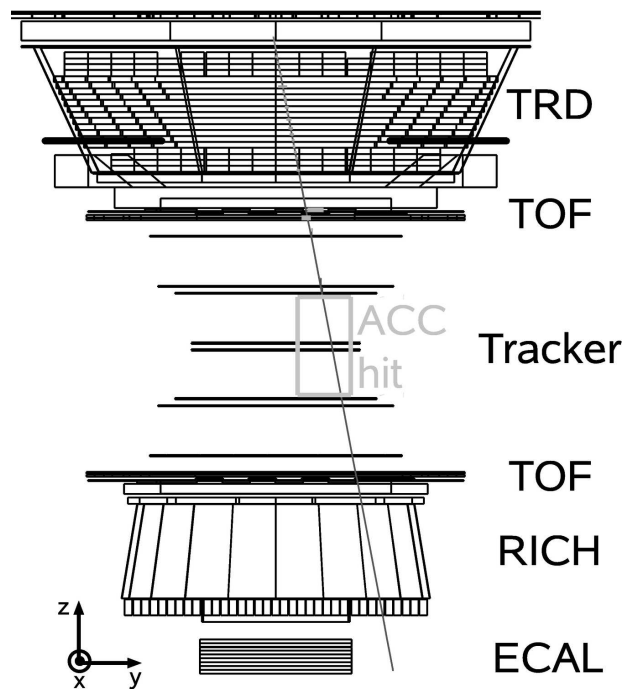}
\caption{Clean event for the ACC analysis during the AMS-02 preintegration runs with atmospheric muons. The outline of the ACC hit shows only a part of a panel.}
\label{f-eventdisplay_1209556604_33077}
\end{figure}

Fig.~\ref{f-eventdisplay_1209556604_33077} shows a typical event used in the following analysis. In addition, the AMS-02 coordinate system is shown. The origin is located at the center of the tracker. The axis of symmetry of the ACC cylinder is the same as the $z$ axis of the AMS-02 coordinate system.

A track fit was developed to find tracks which point to the ACC. The requirement of a reasonable agreement between TRD and tracker track acts as an effective momentum filter such that only high energetic particles without any significant interactions in the TOF or tracker are taken into account for analysis. 

The analysis starts with the TRD track, which is defined by requiring that at least three out of the four upper layers of the TRD, at least ten of the twelve intermediate layers and again at least three of the four lower layers have hits on the track. A hit is considered to be on the track if its horizontal distance $\sqrt{\Delta x^2+\Delta y^2}$ from the track is less than 0.6\,cm (= straw tube diameter).

To select clean single track events the new straight line fit in the tracker requires at least three layers each with only one reconstructed hit within a road around the extrapolated TRD track. In addition, only tracks with reasonable $\chi^2$/dof values are taken into account. For a reliable track fit two additional cuts are applied to accept only clean single track events for the analysis. The ratio of the total number of hits in the TRD to the number of hits on the TRD track has to be smaller than 1.5. In a similar way, the ratio of the total number of reconstructed tracker hits to the number of hits on the fitted track must be smaller than 1.3. These two ratio requirements lower significantly the number of events having additional hits resulting from interactions of the primary particle in the detector. On average, the intersection of the track on the ACC cylinder can be extrapolated with a mean RMS of about 0.7\,mm.  

Furthermore, the comparison of the TRD track and the tracker track showed that TRD and tracker are on there nominal positions with respect to each other and proof the high integration precision. This is done by extrapolating both tracks to the $z$ position of the upper TOF and examining their distance at this position. The resolutions in $x$ and $y$ direction are 0.55\,cm and 0.41\,cm, respectively.

\begin{figure}
\includegraphics[width=1.0\linewidth]{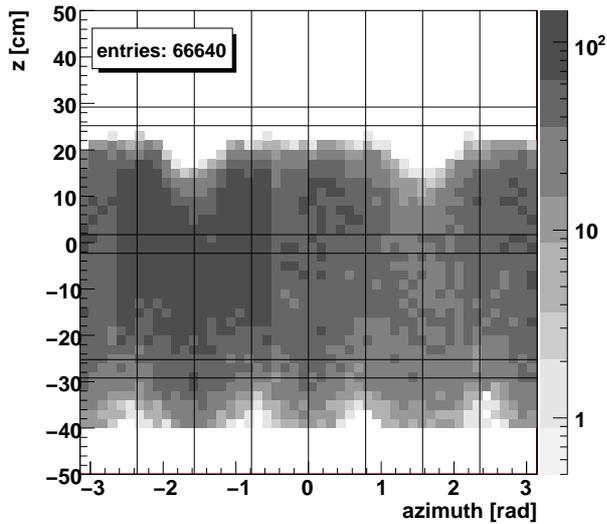}
\caption{Azimuthal and $z$ occupancy on the ACC of fitted tracks surviving all cuts. The vertical lines indicate the positions of the slot regions between two sectors and the horizontal lines the positions of the tracker layers. The color code on the right shows the number of entries.}
\label{f-090216_4_0_6_0_4_2_3_phi_z}
\end{figure}

The ACC cylinder is located between $z=-41.5$\,cm and $z=+41.5$\,cm. The track occupancy distribution in azimuth angle and $z$ on the ACC cylinder shows a structure due to the shape of the TRD, the TOF and the tracker planes (fig.~\ref{f-090216_4_0_6_0_4_2_3_phi_z}). No hits above $z\approx20$\,cm are visible because it was required that the first tracker plane shows a hit on the track.

\begin{figure}
\includegraphics[width=1.0\linewidth]{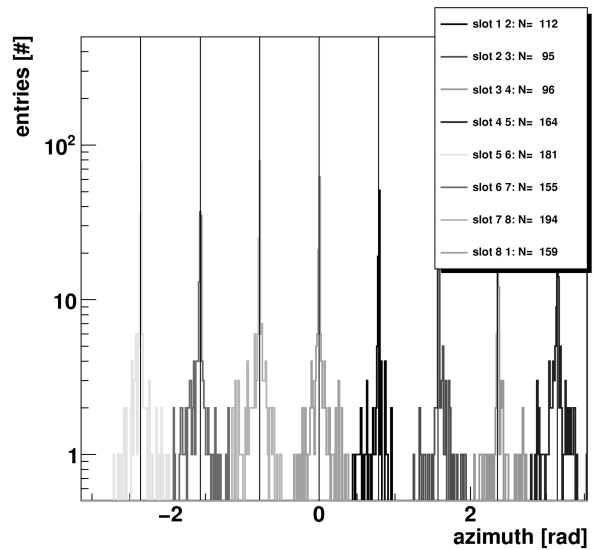}
\caption{Positions of slot regions. The vertical lines indicate the nominal slot positions ($n\cdot\pi/4, (n=-3,\dots,4)$).}
\label{f-090216_4_0_6_0_4_2_3_acc_sector}
\end{figure}

The position of the seven slots between the eight ACC sectors is determined by requiring all four ACC PMTs of adjacent sectors to show clear signals. The frequency distribution of these events is shown as a function of the azimuth angle in fig.~\ref{f-090216_4_0_6_0_4_2_3_acc_sector}. The data shows that the used tracker fit is able to reproduce the geometry of the ACC.

\subsection{Inefficiency Determination}

\begin{figure}
\centering
\includegraphics[width=1.0\linewidth]{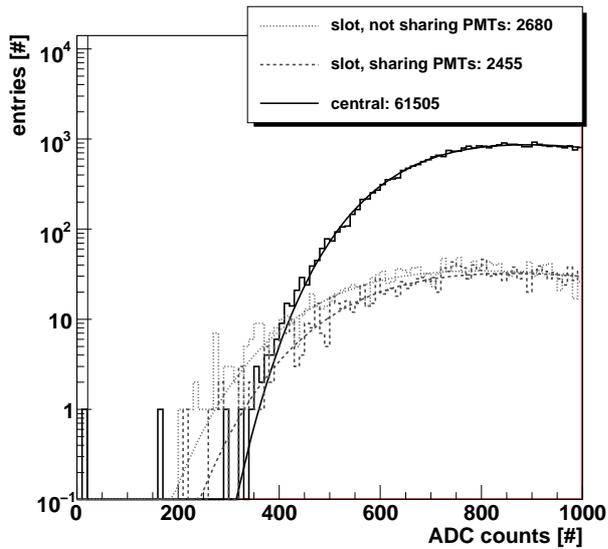}
\caption{Highest ADC values of the complete ACC system for the central region of the ACC panels and the two different slot regions in an interval of 15\,mrad around the slots. The vertical line indicates the cut for the definition of a good event.}
\label{f-090216_4_0_6_0_4_2_3_acc_adc_highest_slot_central}
\end{figure}

In the following, a good ACC event is defined to show at least in one PMT of the sector to which the fitted track is extrapolated an ADC value larger than 21\,ADC counts (average RMS of PMT pedestals: 7\,ADC counts).  ACC events not fulfilling this requirement are called 'missed'. The distribution of the highest ADC values of each event for the central part of a panel, the slot region between two sectors and the slot within a sector is shown in fig.~\ref{f-090216_4_0_6_0_4_2_3_acc_adc_highest_slot_central}. As expected, the slot between two sectors shows the smallest MOP value. Within a sector the mean ADC values as a function of azimuth angle show clear drops at the slot positions (fig.~\ref{f-acc_phi_mod_adc_highest_bothmean}). The mean drop ($\approx20$\,\%) for the slot region between panels sharing their PMTs is not as strong as for the slots between sectors ($\approx30$\,\%). In addition, the spectrum of the central part shows one missed event below the cut of 21\,ADC counts which cannot be explained by statistical fluctuations of the fitted Landau distribution. The mean inefficiency $I$ for the pre-integration of the complete ACC system is calculated to be \begin{equation}I=1.5^{+2.3}_{-1.1}\cdot 10^{-5} < 7.2\cdot10^{-5}\mbox{ (95\,\% confidence level).}\end{equation} This result is beyond the requirement of $10^{-4}$ and dominated by only one missed event which was measured while the complete experiment was rotated around the $y$ axis by -90$^\circ$. By assuming that muons preferbly come from above it probably crossed the ACC from the outside before traversing tracker and TRD. In this case the probability for unclean events due to interactions, e.g. in the tank, is much higher. Therefore, we interpret the derived inefficiency as an upper bound.

\section{Further ACC Properties\label{s-further}}

Besides the inefficiency study the data were used to extract further detector properties. This is important for the development of a realistic signal behavior for the ACC simulation. 

\begin{figure}
\centering
\includegraphics[width=1.0\linewidth]{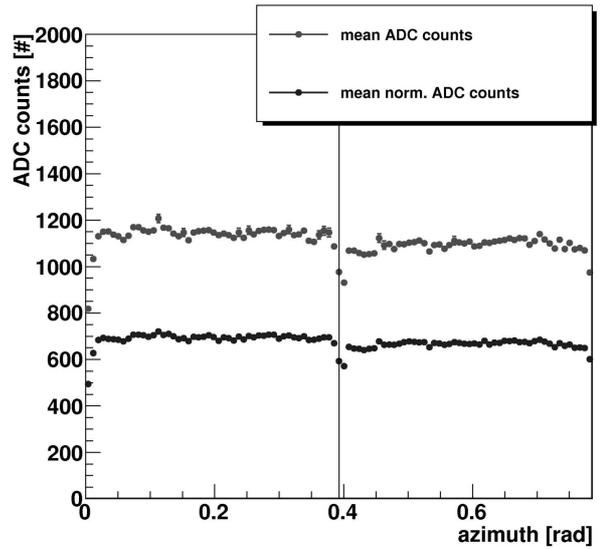}
\caption{Mean (normalized) ADC values vs. the azimuthal position in an ACC sector. The line indicates the slot between the two panels of the sector.}
\label{f-acc_phi_mod_adc_highest_bothmean}
\end{figure}

Knowing the direction and the intersection of a particle with the ACC, the pathlength in the scintillator can be calculated and compared to the measured ADC value. The mean ADC value increases non-linearly with pathlength $l$ and can be parametrized by:
\begin{equation} ADC(l)=(929.9\pm12.4)\cdot\sqrt{l/\mbox{cm}-(0.25\pm0.07)}.\end{equation}
This can be used to normalize all measured charges to the panel thickness which was already used for fig.~\ref{f-acc_phi_mod_adc_highest_bothmean}.

Furthermore, the PMT signal pulseheight decreases due to attenuation if the distance of the particle intersection to the PMT increases.  Additional signal drops at both ends of a panel result from the smaller active to passive material ratio with respect to the rest of the panel due to the method of WLS fiber embedding. A good description of this behavior is achieved by a polynomial fit to the mean normalized signal as a function of position $z$ along the panel:
\begin{equation} \begin{split}ADC(z) &= (-2.5\pm 0.3)\cdot10^{-5}\cdot(z/\mbox{cm})^4\\&-(1.55\pm0.09)\cdot z/\mbox{cm}+(622\pm2).\end{split}\end{equation}The measured signal behavior across (fig.~\ref{f-acc_phi_mod_adc_highest_bothmean}) and along the ACC panels were used for a simulation of the ACC inefficiency during operation in Space. This takes into account that the angular particle distributions of atmospheric muons (mostly from above) and cosmic rays (isotropic) are different which results in longer pathlengths in the scintillator material for an isotropic distribution. Therefore, the average signal height will be larger for cosmic rays. Neglecting the missed event during the pre-integration and extrapolating the Landau fits (fig.~\ref{f-090216_4_0_6_0_4_2_3_acc_adc_highest_slot_central}) to lower signals this simulation led to an ACC detection inefficiency of $I<3.2\cdot10^{-7}$ (95\,\% confidence level). 

\section{Conclusion}

The atmospheric muon analysis with AMS-02 showed that the production of the ACC veto counter was successful.
 
This project is funded by the German Space Agency DLR under contract No. 50OO0501.

\section{Acknowledgments}

The authors wish to thank the following people for their
support: A.~Basili, V.~Bindi, V.~Choutko, E.~Choumilov, A.~Kounine and  A.~Lebedev for the help during data taking at CERN and with the AMS software.


\begin{thebibliography}{99}
\bibitem{ams04} R.~Battiston. "The anti matter spectrometer (AMS-02): a particle physics detector in space," \textit{J. Phys.: Conf. Ser.}, vol. 116, 012001, May 2008.
\bibitem{doe08} Ph.~von Doetinchem \etal, "The AMS-02 Anticoincidence Counter," IPRD08 conference proceeding, submitted to \textit{Nucl. Phys. B - Proc. Sup.}, 2008. Available: arXiv:0811.4314	
\bibitem{doe09} Ph.~von Doetinchem, "Search for Cosmic-Ray Antiparticles with Balloon-borne and Space-borne Experiments," Ph.D. Thesis, RWTH Aachen University, 2009. Available: arXiv:0903.1987	
	     
\end{thebibliography}
\end{document}